\documentclass[aps,preprint,a4paper,showpacs,showkeys,superscriptaddress]{revtex4-1}

\usepackage{latexsym}
\usepackage{amsmath,amssymb}
\usepackage{graphicx}
\usepackage{subfigure}
\usepackage{xcolor}
\usepackage{hyperref}     
\hypersetup{colorlinks,%
  citecolor=blue,%
  linkcolor=cyan,%
  pdftex}

\begin{document}


\title{The first law of thermodynamics in Lifshitz black holes revisited}

\author{Yongwan Gim}
\email[]{yongwan89@sogang.ac.kr}
\affiliation{Department of Physics, Sogang University, Seoul 121-742,
  Republic of Korea}%
\author{Wontae Kim}
\email[]{wtkim@sogang.ac.kr}
\affiliation{Department of Physics, Sogang University, Seoul 121-742,
  Republic of Korea}%
  \affiliation{Research Institute for Basic Science, Sogang University, Seoul, 121-742, Republic of Korea}%
\author{Sang-Heon Yi}
\email[]{shyi@yonsei.ac.kr}
\affiliation{Department of Physics, College of Science, Yonsei University, Seoul 120-749, Korea}

\date{\today}

\begin{abstract}
We obtain the mass expression of the three- and five-dimensional Lifshitz black holes by employing the recently proposed quasilocal formulation of conserved charges, which is based on the off-shell extension of the ADT  formalism. Our result is consistent with the first law of black hole thermodynamics and resolves the reported discrepancy between the ADT formalism and the other conventional methods. The same mass expression of Lifshitz black holes is obtained  by using another quasilocal method by Padmanabhan. 
 We also discuss the reported discrepancy in the context of the extended first law of black hole thermodynamics by allowing the pressure term.
\end{abstract}



\maketitle


\section{Introduction}
\label{sec:intro}

 Recently, there has been much attention to Lifshitz black holes~\cite{Danielsson:2009gi, Mann:2009yx, Bertoldi:2009vn, AyonBeato:2009nh, AyonBeato:2010tm}, because they may give rise to a new perspective on a condensed matter physics
 via the Lif/CFT correspondence which is one of the extension of the AdS/CFT correspondence~\cite{Kachru:2008yh, Maldacena:1997re, Maldacena:1997zz, Hartnoll:2009sz, Son:2008ye}. 
 Besides various applications in the dual field theory context, this type of black holes draw some renewed interests  in  traditional approaches to black hole physics, 
 partly because  Lifshitz black holes have the anisotropic scaling behavior between time and space.  
 In this regard, one may recall that the AdS/CFT correspondence or holography has  shed new light on the traditional approach to a gravity theory. 
  Concretely, conserved charges of black holes, which are identified with corresponding physical quantities of the dual field theory, can be addressed 
  in the context of the holographic renormalization~\cite{Henningson:1998gx, de Boer:1999xf, deBoer:2000cz, Balasubramanian:1999re}. In the light of the power of  the AdS/CFT correspondence,  
  it is strongly anticipated that the holographic approach should give us essentially the same results on conserved charges of black holes as the traditional approach. Since there exist some apparent differences in the formulations, 
   there were some studies on the relation between the holographic and the  traditional approaches to conserved charges~\cite{Papadimitriou:2005ii,Hollands:2005wt,Hollands:2005ya,Fischetti:2012rd}.

 Among the traditional approaches to conserved charges, the Abbott-Deser-Tekin (ADT) method, which is covariant, has been known to produce the completely consistent results with the AdS/CFT correspondence in various cases.  According to the general arguments given in~\cite{Hollands:2005ya}, any consistent covariant method for charges should give us essentially the same results with the holographic approach known as boundary stress tensor method~\cite{Henningson:1998gx,Balasubramanian:1999re}.  Therefore, it is very intriguing if there would be an inconsistency between results from the ADT method and the boundary stress tensor method, since the argument in Ref.~\cite{Hollands:2005ya} depends on the general structure of a gravity theory.  If one could find an example which reveals an inconsistency between those formalisms, one needs to reconsider which steps or which methods break down in such a case. In fact, in a specific higher curvature theory of gravity, it has been known that there is a conflict between the traditional ADT method~\cite{Devecioglu:2010sf} and the boundary stress tensor one~\cite{Hohm:2010jc, Ross:2009ar, Mann:2011hg}. More specifically, the mass expression of Lifshitz black holes in new massive gravity from the traditional ADT method  is claimed to be given by $M_{DS}=7r_H^4/(8G\ell^4)$ where  $r_H=\ell\sqrt{m}$ \cite{Devecioglu:2010sf},
while the one  from other methods  is calculated as 
$M=r_H^4/(4G \ell^4)$ \cite{Gonzalez:2011nz, Hohm:2010jc, Myung:2009up}.

 However, it has not yet been known  that the above conflict is essential   or just superficial, since identifying conserved charges of black holes  becomes rather involved in the case of Lifshitz black holes in a higher curvature gravity. One may guess that the consistency of the boundary stress tensor method  with the Euclidean action approach \cite{Gonzalez:2011nz} and dilaton gravity approach \cite{Myung:2009up}  indicates the mass expression from the boundary stress tensor   method is correct one in this specific case. Furthermore, the claimed expression for the mass of the Lifshitz black hole from the traditional ADT method~\cite{Devecioglu:2010sf}  does not satisfy the first law of black hole thermodynamics 
while the others respect the first law.   In this paper we would like to address this issue and resolve the conflict between the ADT method and the holographic method by using the appropriate adaptation of the traditional ADT method developed in~\cite{Kim:2013cor}.


 In the original ADT method \cite{Abbott:1981ff, Abbott1982, Deser:2002jk, Deser:2002rt, Senturk:2012yi}, which is a covariant generalization of the Arnowit-Deser-Misner(ADM) method~\cite{Arnowitt:1962hi},   the metric is
linearized as $g_{\mu\nu} = \bar{g}_{\mu\nu} + h_{\mu\nu}$ at the asymptotic infinity
where $\bar{g}_{\mu\nu}$ denotes a background metric and $h_{\mu\nu}$ does the fast vanishing  perturbed metric.  This fast falloff condition of the perturbed metric allows us to obtain the finite conserved charges by this linearization. 
However, in the three-dimensional Lifshitz black hole, it is not so clear that the above fast falloff condition on the perturbed metric $h_{\mu\nu}$ is satisfied.  More explicitly,  the metric of the Lifshitz black hole may be taken in the form of 
 $ ds^2=-g_{tt}dt^2+g_{rr}dr^2+r^2d\phi^2 $
with $g_{tt}=r^6/\ell^6-mr^4/\ell^4$ and $g_{rr}=(r^2/\ell^2 - m)^{-1}$.
By taking the background geometry  as the case of $m=0$ in this  metric, one can see that, in the metric component  $g_{rr}$, the background metric $\bar{g}_{rr}=\ell^2/r^2$ becomes exclusively dominant term at the asymptotic infinity, while the perturbed metric $h_{rr}$ vanishes sufficiently fast there. 
On the other hand, though the dominant term of  the metric component $g_{tt}$ at the asymptotic infinity seems to be the background metric $\bar{g}_{tt} = r^6/\ell^6$,  the perturbed metric component $h_{tt} = mr^4/\ell^4$  is also divergent there. So,  it is a  little bit subtle  to regard $h_{tt}$  as the perturbed part  at the asymptotic infinity 
even though their ratio asymptotically vanishes. Consequently, the falloff boundary condition of Lifshitz black holes violates the original assumption in the ADT method. That is to say, the $h_{tt}$ component in this case falls off too slowly  to ensure the validity of the linearized ADT formalism even at the asymptotic infinity. 


There is a quasilocal generalization of the ADT formalism to obtain conserved charges of black holes~\cite{Kim:2013zha}, which can be used even with slow falloff conditions. Furthermore, this formulation allows  us to identify conserved charges in the interior region of black holes not just at the asymptotic infinity in the sense of quasilocal charges. 
Though the off-shell ADT potential as an extension of the original ADT method  was used in the higher derivative theory of gravity for computational convenience~\cite{Bouchareb:2007yx, Nam:2010ub}, it was shown to have more interesting aspects: the off-shell ADT potential is equivalent to the linearized off-shell Noether potential 
up to the surface term~\cite{Kim:2013zha}. This means that the off-shell ADT method can be related directly to the covariant phase space formalism~\cite{Lee:1990nz,Wald:1993nt,Iyer:1994ys,Iyer:1995kg,Wald:1999wa} at the off-shell level.  
 By integrating the ADT potential along the one-parameter path in the
 solution space ({\it i.e.} the on-shell space) \cite{Barnich:2003xg,
   Barnich:2004uw, Barnich:2001jy, Wald:1999wa, Barnich:2007bf}, 
 quasilocal conserved charges can be calculated \cite{Kim:2013zha}. These charges are  consistent with the traditional ADT method at the linearized level and provide us  the consistent non-linear completion of the linearized ADT method. 
As a matter of fact, there is a more practical advantage in this formulation:
 the quasilocal conserved  charges  corresponding to Killing vectors  can be obtained  from the Lagrangian 
without resorting  to the complicated equations of motion even for a higher curvature gravity. By using this formulation or adaptation of the original ADT method, which may be called as the quasilocal ADT method,  we would like to address the above issue on the conserved charges. 
  
There exists another method presented by  Padmanabhan 
in order to derive conserved charges of black holes directly from the relationship 
between gravitational field equations and thermodynamics \cite{Padmanabhan:2012gx}. 
The essential ingredient in this method is to rearrange the equation of motion in order to obtain  
the form of thermodynamic first law of black holes. Since this approach uses the local equation of motion not the integration of a certain potential on the asymptotic space,  conserved charges obtained in this approach can be regarded as the quasilocal quantities. 
In the simplest model  of  static 
spherically symmetric black holes in the presence of matters, the equation of motion
is decomposed into  three parts; the first one corresponds to the mass, the second one 
does the entropy, and the last one does the pressure.  These
eventually yield the mass, entropy, and pressure of black holes when one can identify the black hole 
temperature appropriately. 
In this approach, one may note that
the pressure may have  classical or quantum-mechanical 
origins. Particularly in the latter case, the quantum effects can be incorporated into  equations of motion through the metric function by yielding the semi-classical equations of motion. At first glance, this approach seems to be different from the standard first law of black hole thermodynamics. However, one may notice that
after the pressure term is eliminated appropriately  the mass and entropy can be matched to the conventional ones so that 
they are coincident with the ADM mass and the Wald entropy
\cite{Son:2013eea, Kwon:2013dua}.
In association with thermodynamic phase transition, this pressure term might be relevant. In this context one may note that
there are some attempts on including  pressure-volume type terms in the first law of black hole thermodynamics~\cite{Kubiznak:2012wp, Chen:2013ce, Hendi:2012um, Cai:2013qga}.
 
 In this work, we would like to obtain the quasilocal mass and entropy for the three- and five-dimensional Lifshitz
black holes by using the quasilocal ADT method. The mass expressions of those black holes turn out to be invariant along the radial direction. As a check of our mass expression which is valid even near the horizon, we rederive the identical expression  by using the Padmanabhan's quaislocal method.  In the end we can show that  the first law of black hole thermodynamics or/and boundary stress tensor method are completely consistent with the quasilocal ADT method.
In section \ref{sec:quasi}, we recapitulate the quasilocal formulation of conserved charges in Ref.
 \cite{Kim:2013zha}, which provides a very convenient way 
to determine quasilocal conserved charges of  black hole. 
By applying this formula to the Lifshitz black holes in section \ref{sec:lifshitz},  
we find the quaislocal mass and entropy of the black hole, and check that the ADT method is consistent with the 
first law of black hole thermodynamics and eventually with  boundary stress tensor method \cite{Hohm:2010jc}.
In section \ref{sec:th}, the mass and entropy of the Lifshitz black hole are obtained by the Padmanabhan method 
and the results turn out to be the same with those in section  \ref{sec:lifshitz}. 
Finally,  some discussion will be given  in section \ref{sec:discussion}.
  
\section{Quasilocal formulation of conserved charges}
\label{sec:quasi}

In this section we would like to encapsulate the formulation of quasilocal conserved charges
developed in Ref.~\cite{Kim:2013zha}, which may be regarded as the quasilocal adaption of the traditional ADT method.
Let us consider a variation of action with respect to $g_{\mu\nu}$ for
 a generally covariant theory of gravity  in D-dimensional spacetime, which is given as
\begin{equation}
\label{eq:var}
\delta I[g]= \frac{1}{\kappa}\int d^Dx[\sqrt{-g} \mathcal{G}_{\mu\nu}\delta g^{\mu\nu}+\partial_\mu \Theta^\mu(g;\delta g)], 
\end{equation}
where $\mathcal{G}^{\mu\nu}=0$ is the equation of motion for the metric 
and $\Theta^\mu$ denotes the surface term. 
The transformation of the metric,  under the diffeomorphism $\zeta$,  is $\delta_\zeta g_{\mu\nu}=\nabla_\mu\zeta_\nu+\nabla_\nu\zeta_\mu$, and  
 the corresponding transformation of the
 Lagrangian density, $L$   is given by $\delta_\zeta(L \sqrt{-g})=\partial_\mu(\zeta^\mu\sqrt{-g}L)$.
 By using the Bianchi identity $\nabla_\mu\mathcal{G}^{\mu\nu}=0$, one can derive the identically conserved off-shell Noether current 
$\mathcal{J}^\mu$ from Eq.~(\ref{eq:var}) as
\begin{equation}
  \label{eq:2}
J^\mu(g;\zeta) \equiv \partial_\nu K^{\mu\nu} = 2\sqrt{-g}\mathcal{G}^{\mu\nu}(g)\zeta_\nu +\zeta^\mu\sqrt{-g}L(g)-\Theta^\mu(g;\zeta),
\end{equation}
where $K^{\mu\nu}$ is called as the off-shell Noether potential. 
On the other hand, the on-shell ADT current is defined by $J^\mu=\delta \mathcal{G}^{\mu\nu}\xi_\nu$ 
\cite{Abbott:1981ff, Abbott1982, Deser:2002jk, Deser:2002rt, Senturk:2012yi}, where $\xi_\nu$ 
is a Killing vector and $\delta \mathcal{G}^{\mu\nu}$ denotes the generic variation  of the generalized Einstein tensor. 
This ADT current can be elevated to the off-shell current \cite{Kim:2013zha,Bouchareb:2007yx, Nam:2010ub}  in the form of
\begin{equation}
  \label{eq:3}
  J^\mu_{\text{ADT}} \equiv \nabla_\nu Q^{\mu\nu}_{\text{ADT}} = \delta
  \mathcal{G}^{\mu\nu}\xi_\nu+\mathcal{G}^{\mu \alpha}\delta g_{\alpha
    \nu}\xi^\nu-\frac{1}{2}\xi^\mu\mathcal{G}^{\alpha \beta}\delta
  g_{\alpha \beta}+\frac{1}{2}g^{\alpha\beta}\delta g_{\alpha\beta}\mathcal{G}^\mu_\nu\xi^\nu,
\end{equation}
 where $Q^{\mu\nu}_{\text{ADT}}$ is coined as the off-shell ADT potential. 

Now, it can be shown that the off-shell ADT potential is related 
to the off-shell Noether potential.
To this purpose, the diffeomorphism $\zeta$ is taken as a Killing vector 
$\xi$ in the Noether potential. 
Assuming that the Killing vector is preserved as $\delta \xi^\mu =0$,
one can use the following relation on the surface term \cite{Lee:1990nz,Iyer:1994ys},
 \begin{equation}
 \label{lie}
 \mathcal{L}_\xi \Theta^\mu(g;\delta g)-\delta\Theta^\mu(g;\xi)=0,
 \end{equation}
where $\mathcal{L}_\xi$ represents a Lie derivative along the Killing vector
$\xi$ and
the second term denotes the generic variation of the surface term with respect to the
metric $g_{\mu\nu}$.  This relation combined with the off-shell ADT and Noether  potentials yields 
a key relation for the potentials,
 \begin{equation}
   \label{eq:1}
   \sqrt{-g}Q^{\mu\nu}_{\text{ADT}}(g;\delta g)=\frac{1}{2}\delta K^{\mu\nu}(g;\xi)-\xi^{[\mu}\Theta^{\nu]}(g;\delta g).
 \end{equation}
Then, one can calculate the linearized quasilocal ADT charge by using the ADT potential as 
 \begin{equation}
   \label{eq:4}
   \delta Q(\xi)=\frac{2}{\kappa}\int_{\cal B} d^{D-2}x_{\mu\nu} \sqrt{-g} Q^{\mu\nu}_{\text{ADT}},
 \end{equation}
 where the integration domain ${\cal B}$ does not need to be located at the asymptotic infinity. 
Since  we have adopted the off-shell potential,  one may take a more generic linearization in this formulation than the one   in
the conventional on-shell ADT method.  In the traditional  ADT method used in Ref.~\cite{Devecioglu:2010sf},  the linearization is taken only at the asymptotic infinity under the fast falloff boundary condition which is not satisfied in this case. On the contrary, our linearization is taken along the one-parameter path in the solution space and then  the integration is performed along that path as $Q(\xi)=\int^1_0 ds\,  \delta Q(\xi | s\mathcal{M})$, where 
the free parameter $\mathcal{M}$
is parametrized by the variable $s$ such as  $0 \leq s \mathcal{M} \leq \mathcal{M}$. This linearization is also advocated in other quasilocal formulations~\cite{Wald:1999wa, Barnich:2003xg,  Barnich:2004uw, Barnich:2001jy, Wald:1999wa, Barnich:2007bf}.
By using the relation (\ref{eq:1}) and the formula (\ref{eq:4}) with the one-parameter path integral, the quasilocal conserved charge can be finally written as
 \cite{Kim:2013zha}
\begin{equation}
  \label{eq:5}
  Q(\xi)=\frac{1}{\kappa}\int_{\cal B} d^{D-2}x_{\mu\nu}\Big(\Delta
  K^{\mu\nu}(\xi)-2\xi^{[\mu}\int^1_0ds~ \Theta^{\nu]}(\xi\, |\, s\mathcal{M})\Big),
\end{equation}
where $\Delta K^{\mu\nu}(\xi)$ denotes the finite difference $K^{\mu\nu}_{s=1}(\xi)-K^{\mu\nu}_{s=0}(\xi)$ between the Noether potential  of the black hole solution,  $K^{\mu\nu}_{s=1}(\xi)$  and the one of the vacuum $K^{\mu\nu}_{s=0}(\xi)$. 
The symmetry  given in terms of the Killing vector $\xi$ will determine 
the corresponding charge from Eq.~(\ref{eq:5}).  This formulation can cover the black hole entropy since the entropy is a kind of conserved charge as was shown by Wald~\cite{Wald:1993nt}, which is extended to the case of a theory of gravity with a gravitational Chern-Simons term~\cite{Kim:2013cor} and to the case of the asymptotic Killing vectors~\cite{Hyun:2014kfa}.

\section{Thermodynamic first law in the quasilocal method}
\label{sec:lifshitz}

We are now in a position to present the explicit mass expression of  the three- and five- dimensional  Lifshitz black holes  by employing the quasilocal formulation introduced in 
the previous section. 
 The action for a generic quadratic curvature gravity theory is given by
\begin{equation}
  \label{eq:6}
  I=\int d^Dx \sqrt{-g} \left[\frac{1}{\kappa}(R+2\Lambda)+\alpha R^2+\beta R_{\mu\nu}R^{\mu\nu}+\gamma (R_{\mu\nu\sigma\rho}R^{\mu\nu\sigma\rho} -4R_{\mu\nu}R^{\mu\nu}+R^2) \right].
\end{equation}
The equations of motion for  the above action are  given by \cite{Devecioglu:2010sf}
\begin{equation}\label{eq:eqm}
\mathcal{G}_{\mu\nu} \equiv\frac{1}{\kappa} G_{\mu\nu}+\alpha A_{\mu\nu}+\beta B_{\mu\nu}+\gamma C_{\mu\nu},
\end{equation}
where
\begin{eqnarray}   \label{}
G_{\mu\nu}&=& R_{\mu\nu}-\frac{1}{2}g_{\mu\nu}R-\Lambda g_{\mu\nu},\notag\\
 A_{\mu\nu}&=& 2R R_{\mu\nu}-2 \nabla_\mu\nabla_\nu R+ g_{\mu\nu}(2\nabla_\sigma\nabla^\sigma R-\frac{1}{2}R^2),\notag\\
 B_{\mu\nu}&= &2R_{\mu\rho\nu\sigma}R^{\rho\sigma} - \nabla_\mu\nabla_\nu R + \nabla_\sigma\nabla^\sigma R_{\mu\nu} + \frac{1}{2}g_{\mu\nu}(\nabla_\sigma\nabla^\sigma R- R_{\rho\sigma}R^{\rho\sigma}),\notag\\
 C_{\mu\nu}&=&2R R_{\mu\nu}-4R_{\mu\rho \nu\sigma}R^{\rho\sigma}+2R_{\mu\lambda\rho\sigma} R_\nu^{~\lambda\rho\sigma}-4R_{\mu\rho}R_\nu^{~\rho}-\frac{1}{2}g_{\mu\nu}(R_{\lambda\delta\rho\sigma}R^{\lambda\delta\rho\sigma}-4R_{\rho\sigma}R^{\rho\sigma}+R^2)\,.\notag
\end{eqnarray}
In what follows,  the gravitational constant $\kappa$, the cosmological constant $\Lambda $,
and the other coupling constants $\alpha, \beta, \gamma$ 
will be chosen appropriately according to the specific models taken into consideration.

\subsection{Three-dimensional Lifshitz black hole}
In the three-dimensional  case, various parameters are chosen as 
$\Lambda=13/(2\ell^2),~\alpha=-3\ell^2/4\kappa,~\beta=2\ell^2/\kappa,~\gamma=0$, and $\kappa=16\pi G$.
The Lifshitz black hole solution to the equations of motion (\ref{eq:eqm})  is given by \cite{AyonBeato:2009nh}
\begin{equation}
  \label{eq:9}
  ds^2=-\left(\frac{r^2}{\ell^2}\right)^z\left(1-\frac{m\ell^2}{r^2}\right)dt^2+\frac{1}{\frac{r^2}{\ell^2} \left(1-\frac{m\ell^2}{r^2}\right)}dr^2+r^2d\phi^2,
\end{equation}
where $m$ is a certain integration constant  and the 
dynamical exponent is fixed as $z=3$ in order to satisfy the equations of motion.
In this case, the location of the horizon is given by $r_H=\ell\sqrt{m}$. 
Using the generic formulas for higher curvature terms given in Ref.s~\cite{Kim:2013zha, LopesCardoso:1999cv},
  \begin{align}
    \Theta^\mu(\delta g) &=2\sqrt{-g}[P^{\mu(\alpha \beta)\gamma}\nabla_\gamma\delta g_{\alpha\beta}-\delta g_{\alpha\beta}\nabla_\gamma P^{\mu(\alpha\beta)\gamma}] \label{eq:theta}\,, \qquad  P^{\mu\nu\rho\sigma} \equiv \frac{\partial L }{\partial R_{\mu\nu\rho\sigma}}\,,  \\
    K^{\mu\nu} &=\sqrt{-g}[2P^{\mu\nu\rho\sigma}\nabla_\rho\xi_\sigma -4\xi_\sigma\nabla_\rho P^{\mu\nu\rho\sigma}] \label{eq:K}, 
  \end{align}
and  taking the one-parameter path along the integration constant $m$, it is straightforward to obtain the mass of the Lifshitz black hole. Some detailed steps are as follows. 
By expanding $g_{\mu\nu}$ with respect to an infinitesimal parametrization $m+dm$, 
one can obtain   $\delta g_{\mu\nu}$ and  $\Theta^\mu$ in terms of $m$ and $dm$. Let us take the time-like Killing vector as $\xi_t=(-1,0,0)$ with the appropriate overall sign to avoid the negative mass and the negative entropy.
Now, it is straightforward to compute the Noether potential and the surface term.  After the integration along the one-parameter  path along $dm$, one can obtain
  \begin{equation}
    \label{eq:12}
    \Delta K^{tr}=\frac{8mr^2}{\ell^2}\,, \qquad \int^m_{0} dm~  \Theta^r=-2m^2+\frac{8mr^2}{\ell^2}\,.
  \end{equation}
 Note that the Noether potential for the vacuum solution vanishes, $i.e.$, $K^{tr}_{s=0}=0$.
Finally, it can be shown that the mass $M$ of the Lifshitz black hole is given by
  \begin{align}
    \label{eq:13}
    M &\equiv Q(\xi_t) =\frac{1}{16\pi G}\int^{2\pi}_0d\phi\sqrt{h}\Big[2\epsilon_{tr} \Delta K^{tr} -2\epsilon_{tr}\xi^t\int^1_0ds \Theta^r\Big] \notag \\
&=\frac{r_H^4}{4G \ell^4},
  \end{align}
where $h$ denotes determinant of the induced metric.
 We would like to emphasize that this mass expression is valid even in the interior region of the black hole space time not only at the asymptotic infinity. In fact, our mass expression is invariant along the radial coordinate $r$, which reveals the quasilocal nature of our construction of the ADT charges.

The expression in Eq.~(\ref{eq:13}) at the asymptotic infinity  is coincident with the result which has been 
obtained from the other methods \cite{Hohm:2010jc, Gonzalez:2011nz, Myung:2009up}
but it is different from the claimed expression $M_{DS}=7r_H^4/(8G\ell^4)$ in Ref. \cite{Devecioglu:2010sf}. 

Note that in our approach we have employed the linearization  of parameters in the black hole solutions 
instead of the vacuum solution and integrated such linearized expression along the one-parameter path in the solution space in order to evaluate finite physical quantities. This approach has been advocated in Refs. \cite{Wald:1999wa,Barnich:2001jy, Barnich:2004uw, Kim:2013cor} and different from the prescription adopted in Ref. \cite{Devecioglu:2010sf}.  As was mentioned in the introduction, it is not sufficient, in the case of Lifshitz black holes,  to take the linearization around the vacuum solution for obtaining the finite mass expression of Lifshitz black holes, since the falloff boundary condition in this case violates the validity of the traditional linearized ADT method.

As a side remark, we would like to mention that the entropy from our formulation by using Eq.~(\ref{eq:5}) can be obtained as  
\begin{align}\label{thermo}
 \quad S =\frac{2\pi r_H}{G},
\end{align} 
which is identical with the Wald formula  \cite{Wald:1993nt, Iyer:1995kg}. In fact, this entropy in our quasilocal formulation should be identical with the Wald formula,  because it is shown to be equivalent to the covariant phase space formalism generically~\cite{Kim:2013zha,Kim:2013cor}. The above entropy computation is just the check of our quasilocal construction in this specific Lifshitz black hole case.
By noting that the Hawking temperature is determined as  $T_H = r_H^3/(2\pi \ell^4)$ from the definition of the surface 
gravity,  the black hole mass (\ref{eq:13}) and entropy (\ref{thermo}) satisfy 
the first law of black hole thermodynamics as $dM=T_H dS$.  

\subsection{Five-dimensional Lifshitz black hole}

For our convenience in the five-dimensional black hole, the parameters in the action (\ref{eq:6}) 
are chosen as
$\Lambda=2197/(551\ell^2),~\alpha=-16\ell^2/725,
~\beta=1584\ell^2/13775,~\gamma=2211\ell^2/11020$, and $\kappa=1$.
The metric solution was obtained as \cite{AyonBeato:2010tm}
\begin{equation} 
  \label{eq:5Dsol}
  ds^2=-\left(\frac{r^2}{\ell^2}\right)^z\left(1-\frac{m\ell^{5/2}}{r^{5/2}}\right)dt^2+\frac{1}{\frac{r^2}{\ell^2} \left(1-\frac{m\ell^{5/2}}{r^{5/2}}\right)}dr^2+r^2d\Omega_3^2,
\end{equation}
where $\Omega_3$ is the three-dimensional angular part, $z=2$  for the five-dimensional Lifshitz black hole, and $r_H=\ell m^{2/5}$ denotes the horizon location. 
In this example, the time-like Killing vector is taken as $\xi_t=(1,0,0,0,0)$. 
By using Eqs. ~(\ref{eq:theta}) and (\ref{eq:K}) for higher curvature terms,
the surface term and Nother potential are calculated respectively as
  \begin{align}
    \label{eq:5Dpot}
    \int^m_0dm~ \Theta^r &=\frac{33m}{2755\ell}(382m\ell^3-933\ell^\frac{1}{2}r^\frac{5}{2}),\\
    \qquad \Delta K^{tr}&=\frac{33}{5510\ell^3}(1072r^5+719m^2\ell^5-1866m r^\frac{5}{2}\ell^\frac{5}{2}).
  \end{align}
Since the steps are similar to the three-dimensional case, we just present the final  mass expression for the five-dimensional Lifshitz black hole 
\begin{equation}
\label{5dmass}
M=\frac{297r_H^5}{1102\ell^3}\Omega_3,
\end{equation} 
which is different from the mass  expression 
$M_{DS}=536r_H^5\Omega_3/(2755\ell^6)$ given in Ref. \cite{Devecioglu:2010sf}. 
The entropy can also be read off from our quasilocal formulation as
\begin{equation}
\label{5dentropy}
S=\frac{396\pi r_H^3}{551}\Omega_3\,.
\end{equation}
The Hawking temperature is given by $T_H=5r_H^2/(8\pi\ell^3)$.
 It can be easily shown that
the mass \eqref{5dmass} respects the first law of thermodynamics with the 
entropy \eqref{5dentropy} such as $dM=T_H dS$. 

\section{Thermodynamic first law in Padmanabhan method}
\label{sec:th}
In this section, we will derive the conserved charges of Lifshitz black holes
and study the first law of thermodynamics 
by the use of relation between the thermodynamic first law and equations of motion 
based on the Padmanabhan method \cite{Padmanabhan:2012gx}.  This computation confirms our claim that our quasilocal mass expression of Lifshitz black hole is valid even near the black hole horizon.
In the original work,
the mass, entropy, and pressure can be read off from the equation of motion, in particular, 
the entropy is written as the well-known area law and the pressure 
depends on classical or quantum-mechanical matter. Note that
the action  \eqref{eq:6} consists of two parts; one is the Einstein-Hilbert action with the cosmological constant
and the other is composed of the higher-curvature terms. So, there are largely two options whether these two pieces of action
should be treated as a whole, otherwise the higher-curvature terms should be treated as the independent source
which is of relevance to the pressure term. Now, we will choose the first option because in the absence of
the pressure term it was shown that the mass and entropy were written
 as the ADM mass and the Wald entropy in Einstein gravity \cite{Son:2013eea}. 

\subsection{Three-dimensional Lifshitz black hole}
Let us rewrite the metric (\ref{eq:9}) for $z=3$ for convenience as
\begin{equation}\label{metric2}
ds^2=-\frac{r^4}{\ell^4}f(r) dt^2+\frac{1}{f(r)}dr^2+r^2d\theta^2,
\end{equation}
using the function defined by $f(r) \equiv r^2/\ell^2-m$.
The Hawking temperature of the black hole~(\ref{metric2}) is written as 
\begin{equation}\label{tement2}
T_H=\frac{r_H^2 f'(r_H)}{4\pi \ell^2}.
\end{equation}
 Let us consider the equation of motion of $\mathcal{G}^r_r=0$, then it is 
written at the horizon $r_H$ as
 \begin{align}
 \mathcal{G}^r_r=& \frac{1}{8 r_H^2\ell^2}[-52 r_H^2+4a\ell^2f'(r_H)+12\ell^4f'(r_H)^2 \notag \\
 &-2a\ell^4f'(r_H)f''(r_H)-2a^2\ell^4f'(r_H)f'''(r_H)+a^2\ell^4f'''(r_H)^2] \notag \\
 =& 0. \label{EOM2}
 \end{align}
Note that $f''(r_H)=2/\ell^2$ and $f'''(r_H)=0$, so that the equation of motion ~(\ref{EOM2})
can be factorized as
 \begin{equation}\label{EOM3}
 \mathcal{G}^r_r=\frac{8\ell^2}{r_H^2}\left(\frac{f'(r_H)}{2}+\frac{r_H}{\ell^2}\right)\left(\frac{f'(r_H)}{2}-\frac{r_H}{\ell^2}\right) =0 .
 \end{equation}
The factors such as $8\ell^2/r_H^2$ and $(f'(r_H)/2+r_H/\ell^2)$ are always positive,
what it means is that   
\begin{align}\label{EOM4}
\frac{f'(r_H)}{2}dr_H-\frac{r_H}{\ell^2}dr_H=0, 
\end{align}
after multiplying $dr_H$. 
We want to rewrite this equation in the form of the first law of black hole thermodynamics. 
By taking into account a proper factor, we can get
\begin{align}\label{EOM5}
0=\frac{r_H^2 f'(r_H)}{4\pi\ell^2}d\left(\frac{2\pi r_H}{G}\right)-d\left(\frac{r_H^4}{4G\ell^4}\right).
\end{align}
For a given Hawking temperature (\ref{tement2}), 
it can be shown that Eq.~(\ref{EOM5}) is manifestly written in the form of the first law of black hole thermodynamics 
as $0=T_HdS-dM$. Then, it is natural to identify the mass and entropy of the three-dimensional Lifshitz black hole with $M=r_H^4/(4G\ell^4),~S=2\pi r_H/G$, respectively,
where they are exactly coincident with those in section \ref{sec:lifshitz}.

\subsection{Five-dimensional Lifshitz black hole}
Let us consider the metic (\ref{eq:5Dsol}) for $z=2$ as
\begin{equation}\label{metric3}
ds^2=-\frac{r^2}{\ell^2}g(r) dt^2+\frac{1}{g(r)}dr^2+r^2d\Omega_3^2,
\end{equation}
where the function is defined as $g(r) \equiv r^2/\ell^2-m(\ell/r)^{1/2}$, and 
 the Hawking temperature is given by 
\begin{equation}\label{5Dtem}
T_H=\frac{r_H g'(r_H)}{4\pi\ell}.
\end{equation}
In a similar way to the three-dimensional case,
using $g''(r_H)=2/\ell^2-3m\ell^{1/2}/(4r_{H}^{5/2}),~g'''(r_H)=15m\ell^{1/2}/(7r_{H}^{1/2})$,
the equation of motion $\mathcal{G}^r_r=0$ is written as
\begin{align}
\mathcal{G}^r_r&=-\frac{3}{11020m^{4/5}\ell^2}(2909m^{2/5}+208\ell g'(r_H))(5m^{2/5}-2\ell g'(r_H))\notag\\
&=0.
\end{align}
By multiplying  $dr_H$ with some constants to the latter part of equation of motion
 $5m^{2/5}-2\ell g'(r_H)=0$, one can get 
\begin{equation}\label{5DPad}
0=\frac{r_H g'(r_H)}{4\pi\ell} d\left( \frac{396\pi}{551}r_H^3\Omega_3 \right) - d\left(\frac{297}{1102\ell^3}r_H^5\Omega_3\right).
\end{equation}
Using the temperature (\ref{5Dtem}), the above equation is written in the form of the first law 
of thermodynamics of $0=T_HdS-dM$, and the mass and entropy are easily identified with $M=297r_H^5\Omega_3/(1102\ell^3),~S= 396\pi r_H^3\Omega_3/551$. As expected, they are compatible with the expressions in the previous section \ref{sec:lifshitz}.

\section{Discussion}
\label{sec:discussion}
In this work, we have calculated the mass of the three- and five-dimensional Lifshitz black holes
by using the quasilocal formulation of the conserved charges and obtained 
the quasilocal mass consistent with the first law of thermodynamics,
which has also been confirmed by the Padmanabhan method which uses the relation
between the equations of motion and the first law of black hole thermodynamics.
The advantage for these two methods resides in the fact that those do not resort to the background
vacuum metric, so that the result is naturally independent of the vacuum metric. We have resolved  the discrepancy in the mass expression of Lifshitz black holes
 between the naive  ADT method  \cite{Devecioglu:2010sf} and the other ones by showing that the correct way to incorporate the ADT method is to use one-parameter path in the solution space or in other words to use the nonlinear completion of the linearization. This resolution is completely parallel to the case of warped AdS black holes which also requires  such a nonlinear completion of the naive ADT method~\cite{Bouchareb:2007yx,Kim:2013cor}.
 
 One may say that  the first law of black hole thermodynamics should always hold in our quasilocal formulation of the ADT method, because it is shown to be equivalent to the covariant phase space formalism~\cite{Kim:2013zha,Kim:2013cor} and the first law of black hole thermodynamics is proved to hold in that formalism by Wald~\cite{Wald:1993nt}. However, one needs to be cautious about this statement since there are some assumptions in this proof of the first law of black hole thermodynamics. For instance, we have assumed the smoothness or continuity of one-parameter path in the solution space in order to perform the integration along that path, and we have also assumed the validity of  Stokes' theorem in this formal proof. In addition, there is issue on what kind form of  the first law of  thermodynamics should be used  especially in higher derivative theory of gravity. One of such  modification of the simplest form of the first law of black hole thermodynamics was studied by allowing some chemical potentials in the context of new massive gravity~\cite{Oliva:2009ip}.   In the next paragraph,  another possibility of such modification  will be commented by allowing a pressure term.   Therefore, it may have some meaning to check the first law of black hole thermodynamics explicitly because there was a claim that the ADT method is inconsistent with the first law of thermodynamics~\cite{Devecioglu:2010sf}.  

Though we have resolved  the discrepancy in the mass expression of Lifshitz black holes, it might be intriguing to  discuss the interpretation of the result in \cite{Devecioglu:2010sf} 
in the framework done by Padmanabhan method~\cite{Padmanabhan:2012gx} by which the relationship between gravitational field equations and thermodynamics can be found   in the simplest context.
The Lagrangian is assumed to be two parts; one is the Einstein tensor with the cosmological constant
and the other consists of the higher curvature terms, for instance, 
$L=L_0+L_1$, where $L_0 = R+13/\ell^2$ and $L_1 = 3\ell^2 R^2/4+2\ell^2R_{\mu\nu}R^{\mu\nu}$
especially in three dimensions. Then, the first law of thermodynamics corresponding to 
the equation of motion can be written by reshuffling Eq. \eqref{EOM2} as
 $-d\tilde{M}+T_Hd\tilde{S}=\tilde{P}dV$,
 where $\tilde{M}=13r_H^4/(48G\ell^4)$, 
 $\tilde{S}=\pi r_H/(3G)$,
  $\tilde{P}=-5r_Hf'(r_H)/(24\pi G\ell^2) - r_H^2/(24\pi G\ell^4)$ 
  and $V=\pi r_H^2$. The right hand side of the pressure term $\tilde{P}$
comes from the two higher curvature terms in $L_1$ which play a role of source term  
in gravitational equations based on the original procedure in Ref. \cite{Padmanabhan:2012gx}.
Note that the mass and entropy are not familiar with the conventional ones.  
To overcome this problem, the pressure term  can be eliminated so that it can be split into two
parts and eventually they are absorbed into the mass and entropy, respectively. 
Then, the resulting equation becomes the desired expressions as $dM=T_HdS$ 
where  $M=r_H^4/(4G\ell^4)$ 
and $S =2 \pi r_H/G$,
so that the mass and entropy are the same with Eqs. \eqref{eq:13} and  \eqref{thermo},
respectively. 
In other words, it means that if we allow the pressure term, then the mass and entropy can be changed
according to the way to separate the action.
 Conversely speaking, the form of mass can be written in a different way if
the pressure term is allowed in the first law. 
So, one may ask it is possible to accommodate  $M_{DS}$ if we allow the pressure term. 
Supposing that a certain pressure term exists in the first law of black hole thermodynamics  and combining  the claimed $M_{DS}$ with the Wald entropy, we obtain the pressure as $P=-5r_H^2/(4\pi G \ell^4)$ which is unfortunately incompatible with the above pressure $\tilde{P}$.  This means that it is very hard to accommodate the mass $M_{DS}$ as a conserved charge in these frameworks


\acknowledgments 
We would like to thank M. Eune for exciting discussions.
W. Kim was supported by the National Research Foundation of Korea(NRF) grant funded by the Korea government (MOE) (2010-0008359).
 S.-H.Yi was supported by the National Research Foundation of Korea(NRF) grant funded by the Korea government(MOE) (No.  2012R1A1A2004410).


\end{document}